# Yaw attitudes for BDS-3 IGSO and MEO satellites: estimation, validation and modeling with inter-satellite link observations

Chao Yang[1], Jing Guo[1]*, Qile Zhao[1,2]
(*Corresponding author: jingguo@whu.edu.cn)

1. GNSS Research Center, Wuhan University, 129 Luoyu Road, Wuhan 430079, China
2. Collaborative Innovation Center of Geospatial Technology, 129 Luoyu Road, Wuhan 430079, China

# Abstract

The disclosed satellite metadata as well as previous estimations using Revise Kinematic Precise Point Positioning (RKPPP) approach with L-band data have already demonstrated the continuous yaw steering model used by BDS-3 Medium Earth Orbit (MEO) satellites manufactured by China Academy of Space Technology (CAST) in deep eclipse seasons instead of the orbit normal mode. However, the yaw model has not been validated for MEO satellites manufactured by Shanghai Engineering Center of Microsatellites (SECM), as the horizontal phase center offsets (PCO) approaches zeros, similar for BDS-3 Inclined Geostationary Orbit (IGSO) satellites. In this study, the inter-satellite link (ISL) data were used to estimate the yaw angles of BDS-3 IGSO and MEO satellites with accuracy of around 1.49° to investigate their yaw behaviors, particularly in the deep eclipse seasons. The estimates confirm that the IGSO and MEO satellites from CAST show the similar yaw behaviors, while the SECM MEO satellites do not fully comply with the attitude law published by China Satellite Navigation Office (CSNO). The attitude transition postpones from that predicted by CSNO yaw law, and occurs when the yaw angle is less than 5° and the elevation angle of the Sun above the orbital plane (beta angle) crosses 0°. The transition completes within three minutes with a rate about 0.055°/s. A model is proposed to predict these behaviors, and the ISL residuals return to normal levels, and became more stable in the adjacent of midnight and noon points. Once the yaw models are used.

# Introduction

The construction of the global phase of Beidou satellite navigation system (BDS-3) has been fully completed. The constellation consists of 24 Medium Earth Orbit (MEO) satellites, 3 Geostationary Earth orbit (GEO) satellites, and 3 Inclined Geosynchronous Earth orbit (IGSO) satellites. Ten of the MEO satellites are manufactured by Shanghai Engineering Center for Microsatellites (SECM) of the China Academy of Science, while the rest are developed by China Academy of Space Technology (CAST).

The attitude of the GNSS satellite determines its orientation in space. It is essential for GNSS data analysis as it has impacts on the correction of observation errors and the modeling of non-gravitational perturbations. Following International GNSS Service (IGS) convention, the Z-axis of satellite body frame points to the earth center, the Y-axis is the rotation axis of the solar panel and perpendicular to the direction of satellite to the Sun, and the X-axis completes the right-hand coordinate system and points to the velocity direction (Montenbruck et al. 2015). However, the satellite does not follow the nominal attitude law always. The yaw maneuvers occur near the midnight and noon point of the orbital plane in low beta-angle (the elevation angle of the Sun above the orbital plane) regime, as the required yaw rate exceed the maximum values provided by the momentum wheel of satellite attitude control subsystem. This is termed as the midnight-/noon-turn maneuvers. Moreover, the eclipse maneuver happens when satellites cross the eclipse seasons. It is caused by the output of Sun sensors is essentially zero in the shadow of the Earth (Bar-Sever et al. 1996).

Usually, the attitudes of different satellite blocks behavior differently in the maneuver periods. For GPS BLOCK IIA, the eclipse maneuver occurs when the satellite is in eclipse reasons, while the noon-turn maneuver is performed near the noon point. It has been clearly analyzed by Bar-Sever (1996), and the GPS Yaw Model 95 (GYM95) was proposed. Afterwards, a simplified yaw attitude model for GPS BLOCK II/IIA/IIR satellites have been established by Kouba (2009). By using of the reverse Kinematic Precise Point Positioning (RKPPP) method, the yaw attitudes of GNSS satellites can be derived from the epoch-estimated phase center offsets. With the estimated attitude angles, the attitude laws for GPS BLOCK IIF and GLONASS-M satellites have been established (Dilssner et al. 2011a,b). Similar as RKPPP, the attitude angles of QZSS Michibiki satellites has been derived by estimation of the baseline between Submeter-class Augmentation with Integrity Function and the main navigation antennas, and it clearly demonstrate the attitude switch to orbital normal (ON) mode from yaw-steering (YS) model when the beta-angle is below 20 degrees (Hauschild et al. 2012). For the Galileo In-Orbit-Validation/Full-Orbit-Capacity satellites as well as other QZSS satellites, the nominal yaw attitude models have been released by the European GNSS Service Center and Japanese the Cabin Office, respectively (GSA 2017; Cabinet Office 2017).

Different from GPS, GLONASS, and Galileo satellites, the regional BDS (BDS-2) GEO satellites adopts ON mode, while the YS and ON mode are used by BDS-2 IGSO and MEO satellites. The transformation occurs when the absolute value of the beta-angle is less than 4° (Guo et al. 2013; Dai et al. 2015). As the deficiency of ECOM SRP model on the ON mode, the orbit accuracy degenerates significantly when satellites are in ON mode. Hence, for some BDS-2 and all BDS-3 MEO satellites, the ON mode has been abandoned (Dilssner, 2017; Zhao et al. 2018). Furthermore, the yaw attitude model has been established for BDS-3 MEO satellites from CAST (Dilssner, 2017; Wang et al. 2018). However, the model could not predict the yaw behaviors when the absolute of beta angle is less than 0.1° (Wang et al. 2018). A yaw bias of 0.14° is introduced to predict the features of reverse midnight-turn maneuvers yaw maneuvers (Xia et al. 2019). For BDS-3 SECM MEO satellites, the attitude law was presented firstly by Lin et al. (2018), and confirmed later by the disclosure of BDS satellite metadata. However, it has not been evaluated with the real data, as the horizontal PCOs of the SECM MEO satellites approaches zero, and so dese for BDS-3 IGSO satellites.

Unfortunately, it is not possible to precisely estimate the yaw angles with the RKPPP technique for BDS-3

IGSO and BDS-3 SECM MEO satellites, as the horizontal antenna offsets are close to zero. As the eccentricities of SLR are not, high-rate SLR measurements are used to reconstruct yaw angle profile for Galileo and possible BDS (Dilssner et al. 2020), but the observations are rare. For BDS-3, inter-satellite-link (ISL) data provide promising way for yaw attitude estimation. Hence, the aim of this paper is to analyze the yaw behaviors of BDS-3 satellite, in particularly for IGSO and SECM MEO satellites, by using the estimated attitude angles from ISL data. After description of the yaw laws of BDS-3 satellites, the method for yaw angles estimation with ISL data will be presented. Subsequently, the estimated BDS-3 satellite yaw angles are analyzed with focus on the IGSO and SECM MEO satellites, and a model will be proposed to compensate the deficiency of the yaw model for SECM satellites. Following the investigation of the impacts of attitude on ISL residuals, the study is summarized finally.

# BDS-3 yaw models

## CAST IGSO and MEO satellites

By using the estimated yaw profile of BDS-2 I06, the yaw law for BDS-3 CAST IGSO and MEO is proposed by Wang et al. (2018) as follows,

$$\psi(t) = 90° \cdot SIGN(1,\psi_s) + [\psi_s - 90° \cdot SIGN(1,\psi_s)] \cdot \cos\left(\frac{2\pi}{t_{max}} \cdot (t - t_s)\right) \quad (1)$$

where $SIGN(a,b)$ is a sign function in Fortran, which means to take the $a$ value and the sign of $b$. $t_{max}$ is a constant that represents the duration of yaw maneuvers, and $t_s$ represents the start epoch of the yaw maneuvers with the nominal yaw angle $\psi_s$ following:

$$\psi = ATAN2(-\tan\beta, \sin\mu) \quad (2)$$

where $\beta$ is the elevation angle of the Sun above the orbital plane, and $\mu$ is the orbital angle (the geocentric angle between the satellite and the midnight point in the orbital plane). For $\psi_s$ computation, the Sun elevation and orbital angle at $t_s$ epoch, i.e., $\beta_s$ and $\mu_s$, are used. Usually, the satellite experiences the yaw maneuvers with $|\beta| \leq 3°$, and the midnight and noon maneuvers occur when orbit angle $\mu_s$ falls in $[-6°, +6°]$ as well as $[174°, 186°]$, respectively. The corresponding $t_{max}$ approximates 3090 s and 5740s for MEO and IGSO, respectively. It is indicated as WHU model hereafter.

## SECM MEO satellites

According to the document "Beidou/Global Navigation Satellite System (GNSS) Satellite High-precision Application Parameters Definition and Description" (CSNO et al. 2019), BDS-3 SECM MEO satellites also obey the continuous yaw-steering law. Generally, the yaw attitude with β=±3° is applied for the satellite when |β| is less than 3°, expressed as follows,

$$\psi(t) = ATAN2(-\tan\beta_0, \sin\mu) \qquad 0 \leq \beta \leq \beta_0 \quad (3)$$
$$\psi(t) = ATAN2(+\tan\beta_0, \sin\mu) \qquad 0 > \beta \geq -\beta_0$$

where $\beta_0$ equals 3°. When the Sun crosses the orbital plane, the sign of the yaw attitude will be changed to be consisted with that of β angle. It is indicated as CSNO model hereafter.

## Yaw attitude estimated with ISL measurements

The RKPPP technique has been successfully used to determine the attitude behavior for GNSS satellites by estimation of the epoch-wise horizontal PCO of microwave signals. Similarly, the epoch-wise horizontal eccentrical offsets of ISL antenna can be estimated with ISL data to derive the yaw attitude profile as described below.

The discrepancy of actual yaw attitude to the nominal one in epoch t $\Delta\psi(t)$ results in the observed PCO $(x(t), y(t), z(t))$ as follows,

$$x(t) = x_0 \cos\Delta\psi(t) - y_0 \sin\Delta\psi(t)$$
$$y(t) = x_0 \sin\Delta\psi(t) + y_0 \cos\Delta\psi(t) \tag{4}$$
$$z(t) = z_0$$

where $(x_0, y_0, z_0)$ are the nominal PCO. With the $(x(t), y(t), z(t))$ estimated with ISL data, the yaw angle deviation can be obtained as,

$$\begin{bmatrix} \cos\Delta\psi(t) \\ \sin\Delta\psi(t) \end{bmatrix} = \begin{bmatrix} x_0 & -y_0 \\ y_0 & x_0 \end{bmatrix}^{-1} \begin{bmatrix} x(t) \\ y(t) \end{bmatrix} \tag{5}$$

$$\Delta\psi(t) = ATAN2(\sin\Delta\psi(t), \cos\Delta\psi(t))$$

where ATAN2(b, c) is the FORTRAN function for arctan(b, c), which gives a signed angle in the range [−180°, +180°].

Obviously, the accuracy of the estimated yaw attitude is related to the magnitude of the nominal horizontal PCO. The larger the PCO, the better the accuracy. The L-band PCO of the SECM satellites are about 5 cm and 0 cm in the X-axis and Y-axis, hence, reliable attitude estimation is difficult, where the ISL makes it possible due to up to meter horizontal offsets. In this study, we uses the clock-free ISL observations to estimate the yaw attitudes, and it is expressed as,

$$\frac{p_{AB}(t) + p_{BA}(t)}{2} = |(\vec{r}_B(t) + E_B(t) \cdot \delta\vec{r}_B(t)) - (\vec{r}_A(t) + E_A(t) \cdot \delta\vec{r}_A(t))| + c(\delta_A + \delta_B) + \frac{\varepsilon_1 + \varepsilon_2}{2} \tag{6}$$

where $p_{AB}(t)$ and $p_{BA}(t)$ are the forward and backward one-way ISL measurement at epoch $t$; $\vec{r}_A(t)$ and $\vec{r}_B(t)$ are the center-of-mass position of satellite $A$ and $B$ with eccentricity of $\delta\vec{r}_A(t)$ and $\delta\vec{r}_B(t)$ to be estimated; $E_B(t)$ and $E_A(t)$ are the rotation matrix from satellite body frame to the inertial frame. $\varepsilon_1$ and $\varepsilon_2$ are the noises of one-way ISL measurement; $\delta_A$ and $\delta_B$ present the signals hardware delays for each satellites. Similar as RKPPP, the epoch-wise eccentricity of ISL antenna can be estimated with Equation (6), afterwards, the yaw can be estimated with Equation (5).

## Data procession

To study the actual yaw-attitude behavior of BDS-3 IGSO and MEO satellites during eclipse seasons, we have estimated their horizontal eccentricity of ISL antenna epoch-by-epoch using 60s clock-free combinations of ISL data with Position and Navigation Data Analyst (PANDA) software. As the time slot is 3s for forward and backward one-way ISL measurements, there are 20 observables at most for a certain BDS-3 satellite contributing to the eccentricity estimates, but the actually number is lower than 20 due to discontinuous connection as well as data gap.

As orbit determination with ISL, we process the combined clock-free ISL observables directly. For POD, the ISL eccentricity are usually estimated with orbit parameters and satellite-dependent hardware delay parameters. However, Xie et al. (2020) identifies that the post-fit residuals of ISL clock-free observables for some satellite pairs have constant biases, when the satellite-dependent transmit and receiver hardware delays are estimated. Their analysis confirms the biases are link dependent. Hence, the link-dependent biases are calibrated in this study. However, due to the high mathematical correlations among eccentricity of ISL antenna and certain orbital elements, estimating all parameters in one run is unreliable. We therefore decided to keep the eccentricity of ISL antenna fixed to their nominal values for POD to solve the remaining parameters first. Afterwards, we estimate the eccentricity of ISL antenna epoch-by-epoch (with respect to the a priori orbit trajectory) whilst keeping all other parameters fixed. The nominal yaw-attitude model given by Eq. (2) thereby serves as an initial yaw-attitude model. The estimated horizontal eccentricities $x(t)$ and $y(t)$ are substituted into Eq.(5) in order to reveal the yaw bias $\Delta\psi(t)$ between nominal and actual yaw angle. Estimates for the actual yaw angle $\psi(t)$ are finally obtained by sum of the nominal as well as the corrected.

Fig. 1 shows the variation of β angle for each orbital plane from Day of Year (DOY) 100, 2019 to 180, 2020, which is the selected study period in this paper. Generally, BDS-3 MEO satellites are distributed in three orbital planes, i.e., Slot-A (C27-C30, C34, C35, C43, C44), Slot-B (C19-C22, C32, C33, C41, C42), and Slot-C (C23-C26, C36, C37, C45, and C46), whereas the BDS-3 IGSO satellites also orbit in three planes, i.e., Plane I (C39), Plane II (C40), and Plane III (C38). The BDS-3 IGSO satellites will cross eclipse seasons for |β|<9°, while, for MEO satellites, the orbital planes are partially eclipsed for |β| < 13.2°.

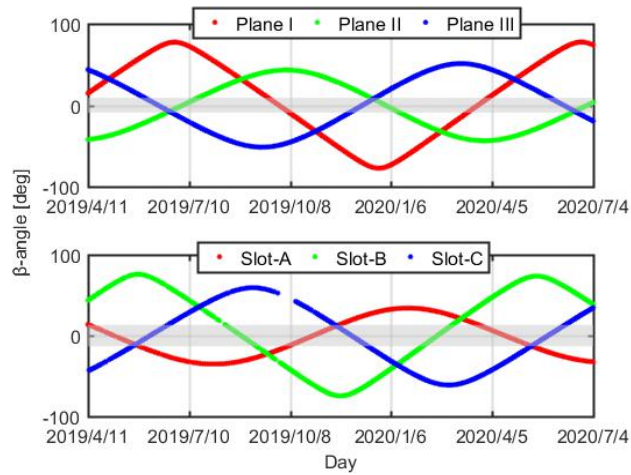

Fig. 1 Elevation of the Sun (β angle) with respect to the orbital planes for BDS-3 IGSO (top) and MEO (bottom) satellites from DOY 100, 2019 to 180, 2020. The shaded area indicates the eclipse season.

# Result analysis

## Estimation accuracy

To verify the accuracy of the estimated yaw angles based on the ISL, the measured ISL data is used to estimate the yaw attitude of the BDS-3 satellite in the non-eclipse season. During this period, BDS-3 satellites can maintain the nominal yaw attitude without maneuvers. Hence, the nominal yaw attitude can be used as the truth to assess the accuracy of the estimated yaw angles. Figure 2 shows the nominal yaw angles (red) and the estimated yaw angles (blue) of the C19 satellite on DOY 230, 2019. The β angle varies from 7° to 6°. It can be seen that the estimated yaw attitudes of the C19 satellite are very closed to the nominal in the entire 24-hour period. The RMS of the estimation errors is about 1.49°. This indicates that a precise yaw attitude estimation can be obtained based on the ISL observations. Hence, ISL measurements can be used to analyze and verify the attitude control modes of the BDS-3 satellites.

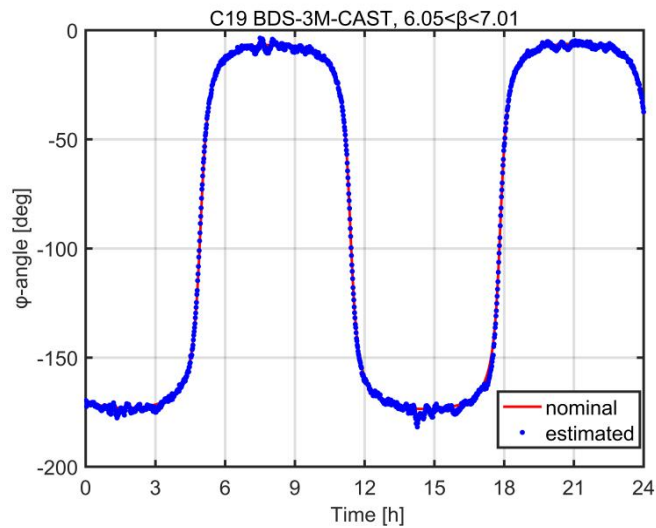

Fig. 2 The nominal (red) and ISL estimated (blue) yaw angles for C19 on DOY 230, 2019 with β of 7°.

## CAST MEO satellites

Figure 3 and Figure 4 illustrates the nominal, estimated and WHU modeled yaw profile of C22 around the midnight as well as noon point during the period of DOY 234 to 240, 2019, when the satellite is in the deep eclipse seasons. It can be clearly observed that behavior of yaw attitude maneuvers near the midnight and noon point is similar, and the satellite starts and finishes maneuvers at fixed orbital angles, i.e., approximate $\pm 6°$ for midnight maneuver, and $180 \pm 6°$ for the noon maneuver. The actual yaw angle of the satellite equals to the nominal at the midnight and noon point to ensure the deviation between the actual and nominal yaw angle as small as possible and thus to reduce the impact of yaw attitude on data analysis. Most importantly, the yaw law expressed by Equ. (1) precisely predicts the yaw behaviors of C22, and the RMS of the differences between the predicted and estimated yaw angles value is 3.18°. Similar behaviors have also identified for other CAST MEO satellites. This indicates that the yaw law expressed by Equ. (1) has good ability to predict the yaw angles of CAST MEO satellites.

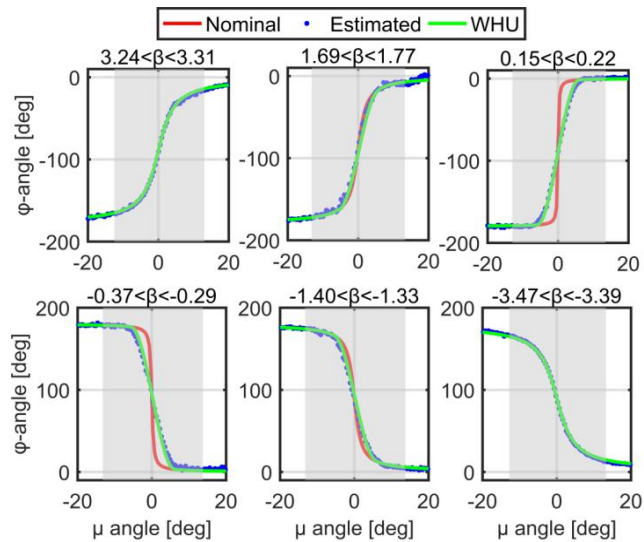

Fig. 3 The nominal (red), estimated (blue), and WHU modeled (green) yaw angles of C22 around the midnight point in the period DOY 234 to 240, 2019.

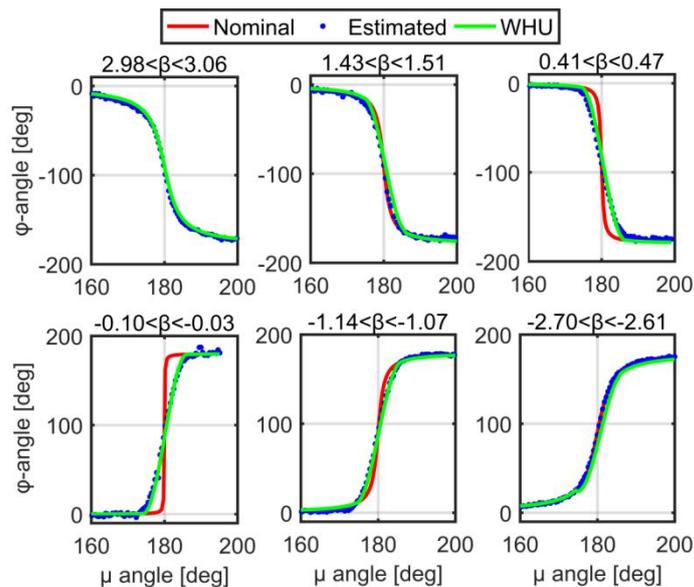

Fig. 4 The nominal (red), estimated (blue), and WHU modeled (green) yaw angles of C22 around the noon point in the period DOY 234 to 240, 2019.

## CAST IGSO satellites

Similarly, Figure 5 and Figure 6 shows the nominal, estimated and WHU modeled yaw angles during the midnight and noon yaw maneuvers for BDS-3 IGSO C38 satellite during DOY 156 to 166, 2020, respectively. It confirms that the spacecraft has obvious maneuvers at the small $|\beta|$. Moreover, it can be noticed that the C38 satellite, during the noon or midnight maneuvers, behaves to a certain extent like the MEO satellites manufactured by CAST. The satellite starts and finishes maneuvers at fixed orbital angles, and the actual yaw angle of the satellite equals to the nominal at the midnight and noon point. However, about 50 min is required to finish the maneuvers for IGSO satellite in the deepest eclipse season, whereas it takes about half time for MEO satellites. This estimation clearly confirm that the yaw law expressed by Equ. (1) can also precisely predicts the yaw behaviors of C38 with accuracy of 4.01°.

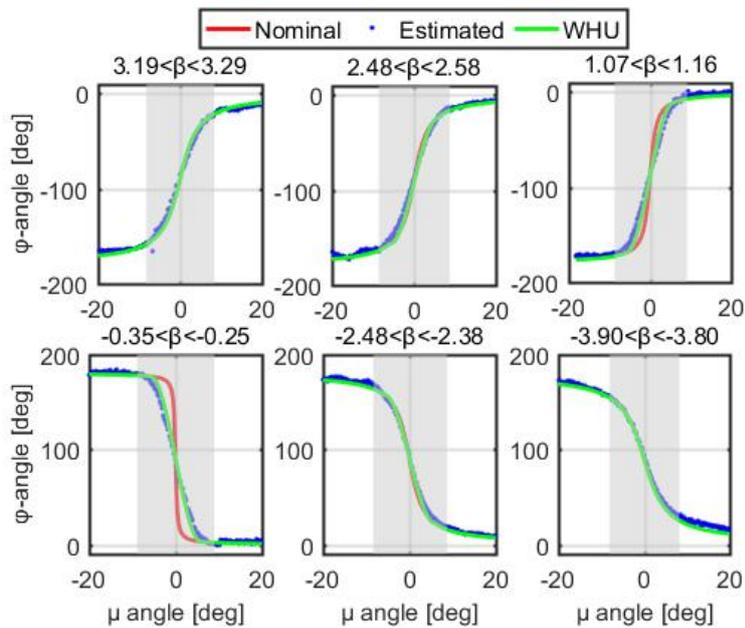

Fig. 5 The nominal (red), estimated (blue), and WHU modeled (green) yaw angles of C38 around the midnight point in the period DOY 156 to 166, 2020.

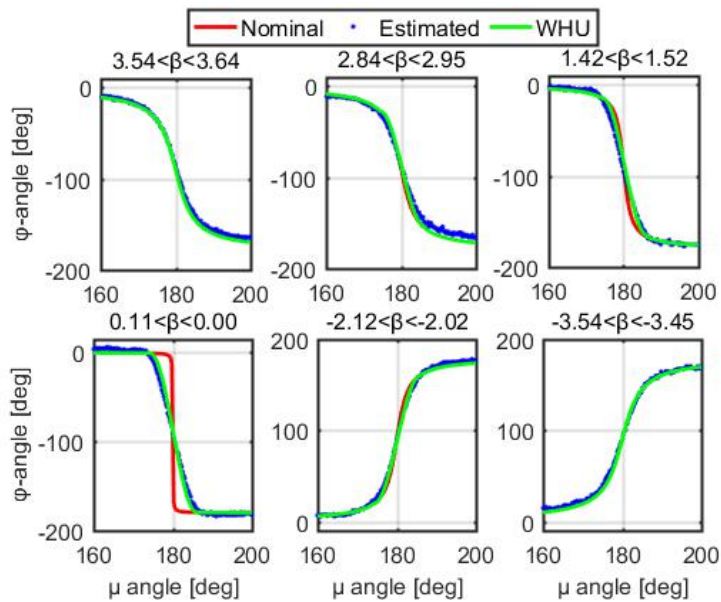

Fig. 6 The nominal (red), estimated (blue), and WHU modeled (green) yaw angles of C38 around the noon point in the period DOY 156 to 166, 2020.

Previously, Xia et al. (2019) observes that the BDS-2 IGSO and MEO satellites with continuous yaw-steering mode will reverse the turn direction for $|\beta| < 0.14°$, hence, a yaw bias of 0.14° is introduced to predict the features. We carefully investigate the estimated yaw angles, and no reversal yaw direction is identified for all BDS-3 IGSO and MEO satellites from CAST, as an example, Fig.7. show the estimated and nominal yaw profile for C24 when the Sun crosses the orbital plane in DOY 156 and 335, 2019, respectively.

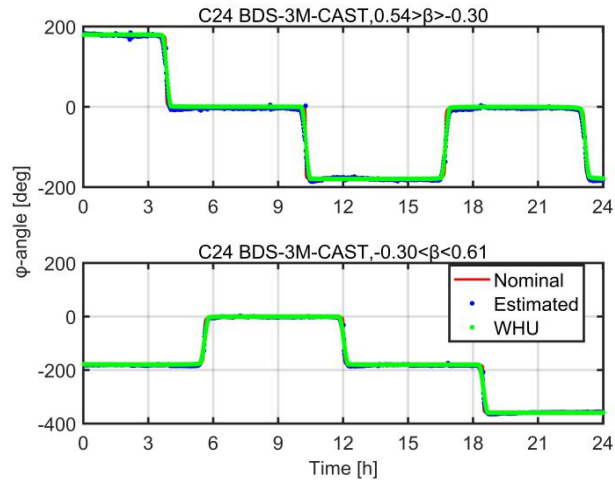

Fig. 7 The nominal (red), estimated (blue) and WHU modeled (green) yaw angles for C24 at the small $\beta$ angles on DOY 156, 2019 and DOY 335, 2019, respectively. The directions of yaw maneuvers are same, and no reversal yaw direction is identified.

## SECM MEO satellites

Furthermore, the yaw angles for SECM MEO satellites are also estimated with ISL data. Figure 8 and Figure 9 show the evolution of the nominal (red), estimated (blue), and CSNO modeled (green) yaw angles for the selected C27 satellite during the deep eclipse seasons (DOY 120 to 132, 2019). It can be clearly observed that the estimated yaw angle equals $\pm 90°$ at midnight and noon point, while the nominal has same value. The actual yaw angles show slight differences with respect to that of the nominal even in non-eclipse seasons, and smaller the absolute of $|\beta|$ angle, greater the differences. This bias disappears until $\beta > 3°$, when the nominal and estimated yaw angle are almost identical. Fortunately, the yaw law expressed in Equ. (3) can reproduce the yaw behaviors quite well. Statistically, the maximum and RMS values of the difference between the predicted yaw angles and estimated yaw angles are 8.5° and 2.81°, respectively.

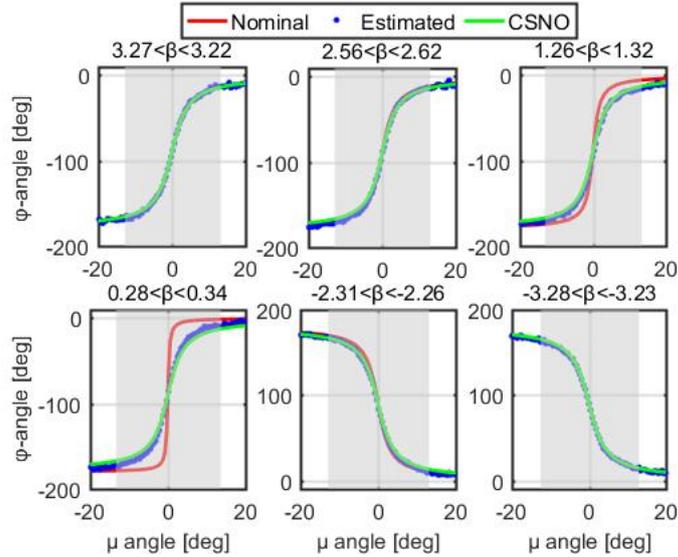

Fig. 8 The nominal (red), estimated (blue), and CSNO modeled (green) yaw angles of C27 around the midnight point in the period from DOY 120 to 132, 2019.

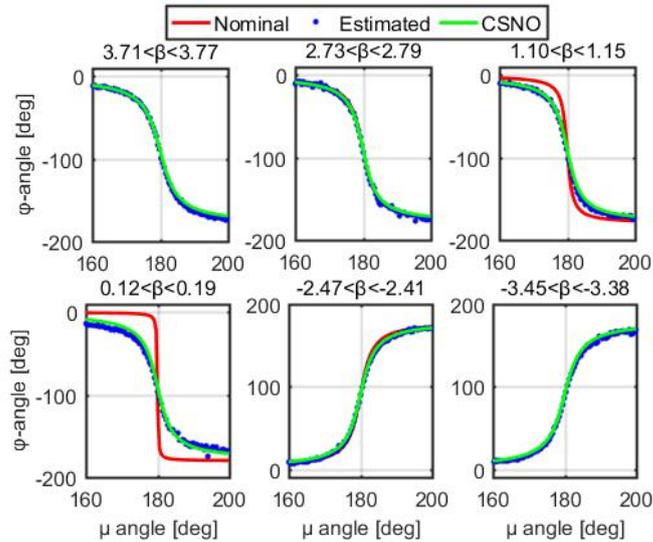

Fig. 9 The nominal (red), estimated (blue), and CSNO modeled (green) yaw angles of C27 around the noon point in the period from DOY 120 to 132, 2019.

However, some abnormal reversal events are observed, shown in Fig.10 for C28 on DOY 304, 2019. For SECM C28, during the period, the Sun crosses the orbital plane from the below to the above. According to Equ (3), a noticeable yaw transition can be expected due to the change of sign of $\beta$ angle. However, it is not true as illuminated by the estimated angles. It seems that the transition shown by the noticeable jump occurs with several hours delay. Hence, the yaw law will be kept as previous, resulting in the reversal direction of yaw maneuvers around the midnight point around 18 h. Same behaviors have been identified for other SECM satellites, but not shown here for brevity.

Table 1 lists the information of ten SECM MEO satellites when the yaw jump occurs. In general, not all of yaw attitude transition occurs immediately when the $\beta$ angle switches the sign. However, all the transition occurs at almost same orbital angle, i.e., around $36.7°$, or same absolute value of yaw angle, i.e., $|5°|$. In this study, we use the yaw angle instead of orbital angle as the condition to predict the yaw transition epoch. The estimate yaw angles show that the jumps occur immediately when the yaw angle less than $|5°|$ and the Sun cross the orbital plane. Hence, the yaw behaviors can be described as follows,

*The yaw attitudes of SECM MEO obey Equ (3), and the transition happens when the $\beta$ passes $0°$ and the absolute of yaw angle is less than $5°$.*

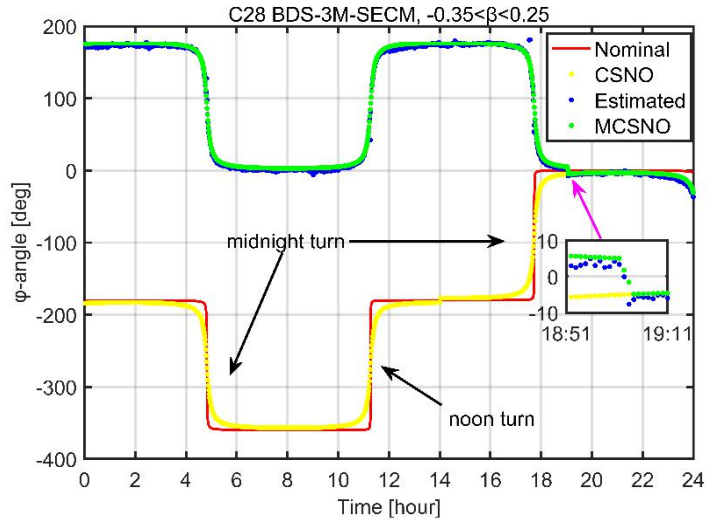

Fig. 10 The nominal (red), estimated (blue), CSNO (yellow) and modified CSNO (MCSNO) (green) modeled yaw attitudes for C28 on DOY 304, 2019.

To investigate the behaviors of SECM satellites during the attitude transition, we re-estimate the yaw angles with 20s ISL data. It seems that the yaw angles are adjusted linearly towards the nominal with fixed rate, which has been estimated for six SECM satellites and listed in Table 2. This transition behaviors can be omitted for modeling, as the maximum bias is around 10°, which has limited impacts on data analysis as well as orbit modeling. All the estimated rates are within $0.1°/s$, and smaller compared to the yaw rate of GPS. As the discrepancy of real and target yaw angles is less than 10°, the yaw maneuver can be completed within 3 minutes. For modeling, the average value of about $0.055°/s$ is used. As plotted in Fig. 10, the new model indicated as Modified CSNO (MCSNO) can predicted the actual yaw behaviors well.

Table 1 The attitude information of all SECM satellites passing through the region where $\beta=0$.

| PRN | DOY | β | μ | φ |
|---|---|---|---|---|
| C29 | 124,2019 | -0.18 | 36.86, | 4.99 |
| C30 | 124,2019 | -0.11 | 36.81 | 5.00 |
| C34 | 124,2019 | -0.05 | 36.90 | 5.00 |
| C27 | 126,2019 | -0.04 | 36.80 | 5.00 |
| C28 | 304,2019 | 0.12 | 36.84 | -5.00 |
| C35 | 302,2019 | 0.15 | 36.88 | -4.99 |
| C25 | 334,2019 | -0.28 | 36.72 | 5.00 |
| C26 | 334,2019 | -0.14 | 36.82 | 5.00 |
| C43 | 107,2020 | -0.13 | 36.61 | 5.02 |
| C44 | 107,2020 | -0.07 | 36.66 | 5.02 |

| C27 | 304,2019 | 0.16  | 36.88  | -4.99 |
| C25 | 155,2019 | 0.09  | 36.77  | -5.00 |
| C29 | 301,2019 | 0.02  | 36.69  | -5.01 |
| C28 | 126,2019 | -0.01 | 36.78  | 5.00  |
| C26 | 155,2019 | -0.00 | 66.62  | 3.26  |
| C30 | 301,2019 | -0.00 | 90.81  | 3.00  |
| C34 | 302,2019 | -0.00 | 138.45 | 4.52  |
| C35 | 124,2019 | 0.00  | 58.29  | -3.53 |

Table 2 The linear fitting of the attitude transition behaviors for four SECM satellites.

| PRN | Fitting yaw angle rate (°/s) |
|-----|------------------------------|
| C27 | 0.067 |
| C28 | 0.062 |
| C29 | 0.049 |
| C30 | 0.032 |
| C34 | 0.075 |
| C35 | 0.045 |

# Validation

The mismodeling of the yaw attitude will lead to correction errors of eccentricities of antenna. If the eccentricities are not calibrated epoch, the mismodeling errors cannot be observed by other parameters, but is only left to the residuals. Hence, the variations of residuals can be used to assess the validation of yaw models. As the horizontal offsets approach zero for L-band antennas, the misorientation cannot be captured by microwave residuals. Hence, ISL observations of CAST and SECM satellites during yaw maneuvers are used for validation. To achieve this, two orbit solutions are determined based on ISL measurements with considering the yaw maneuvers or not. Figure 12 illustrates the ISL residuals of the two solutions for C29 and C34 in the same orbit plane as well as C29 and C21 located in different orbit plane at DOY124, 2019, when C29 and C34 experiences the yaw maneuvers. It can be clearly observed that the ISL residuals near midnight or noon points considerable increase when the nominal attitude is adopted during yaw maneuvers, and the maximum errors can up to 90 cm. By using the yaw attitude models, the errors are significantly reduced, and the magnitudes of the ISL residuals inside and outside the yaw maneuver period are consistently comparable. Additionally, it can be observed that there are three and six jumps for C29 and C21 as well as C29 and C34, respectively. As both C29 and C34 orbits the earth nearly twice with a day, and experience midnight two times and noon one time, whereas the C21 satellite in the another orbital plane do not experiences yaw maneuvers. In addition, the wrong attitude direction will be reflected in the residuals. It is obvious that the reverse attitude causes a maximum error of close to 200 m. The MCSNO corrects this error well and maintains the smoothness of the residuals, which proves that modified model has a strong ability to reproduce the actual yaw attitude.

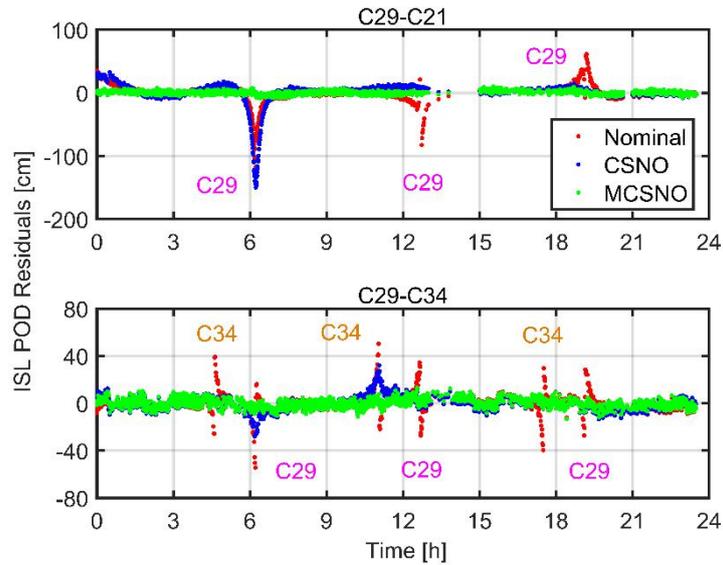

Fig. 11 The out-plane link (C29-C21, top) and in-plane (C29-C34, bottom) link residuals with the nominal attitude (red dots), the CSNO (blue dots), and the modified CSNO model (green dots) in DOY124,2019

# Conclusions and discussions

Based on the ISL observations of BDS-3, the epoch-wise yaw angles are estimated with RKPPP approach. Generally, the estimated yaw angles are in excellent agreement with the nominal yaw angles with accuracy of around 1.5 degrees outside of eclipse reasons. And those clearly show that yaw maneuvers occurs when beta angle are less than 3° for IGSO and MEO satellites from CAST and SECM. For CAST IGSO satellites, the midnight and noon maneuver are similar as that of MEO CAST satellites, and can be precisely predicted by the model of Wang et al. (2018). However, for SECM MEO satellites, their behaviors are a bit different than that of CSNO released model. The transition of yaw attitude when the Sun cross through the orbital plane show a few hours delay, resulting the opposite yaw directions between the modeled and estimated yaw angles when $\beta$ is at a small region. By analysis the estimated angle of all SECM MEO satellites, we identify that the yaw transition occurs when the beta passes 0° and the yaw angle is less than 5°. Hence, the CSNO model has been modified to consider the features. Besides, a linear model is proposed to account for the linear transition of yaw attitude. With the estimated yaw angles, the validation of the proposed and disclosed attitude models is performed. The reversed attitude can be accurately predicted by the new model. And the SECM satellites attitudes adjustments are smoother. Besides, the residuals of Ka-band show that employing the modified yaw model can reduce the residual errors to a normal level.

# Acknowledgements

The CSNO, CAST, SECM are greatly acknowledged for disclosing the satellite metadata as well as ISL data (not public). This work is sponsored by the National Natural Science Foundation of China (41974035;

# References

Bar-Sever Y (1996) A new model for GPS yaw attitude. Journal of Geodesy 70(11):714-723. https ://doi.org/10.1007/BF008 67149

Cabinet Office (2017) QZS-1 Satellite Information. Tech. Rep. SPI_QZS1 Government of Japan, National Space Policy Secretariat. http://qzss.go.jp/en/technical/qzssinfo/index.html

CSNO (2019b) Definitions and descriptions of BDS/GNSS satellite parameters for high precision application. http://www.beidou.gov.cn/yw/gfgg/201912/W020200323534413026471.doc. Accessed on 1 September 2021

Dai X, Ge M, Lou Y, Shi C, Wickert J, Schuh H (2015) Estimating the yaw-attitude of BDS IGSO and MEO satellites. Journal of Geodesy 89 (10):1005–1018. https://doi.org/10.1007/s00190-015-0829-x

Dilssner F, Springer T, Enderle W (2011a) GPS IIF yaw attitude control during eclipse season. AGU Fall Meeting, San Francisco.http://acc.igs.org/orbits/yaw-IIF_ESOC_agu11.pdf. Accessed 9 Dec 2011

Dilssner F, Springer T, Gienger G, Dow J (2011b) The GLONASS-M satellite yaw-attitude model. Advance in Space Research 47(1):160-171. https://doi.org/10.1016/j.asr.2010.09.007

Dilssner F (2017) A note on the yaw attitude modeling of BeiDou IGSO-6, a report dated November 20,2017.http://navigation-office.esa.int/attachments_24576369_1_BeiDou_IGSO-6_Yaw_Modeling.pdf. Accessed 21 Jan 2018

Dilssner F, Schönemann E, Mayer V, Springer T, Gonzalez F, Enderle W (2020) Recent advances in Galileo and BeiDou precise orbit determination at ESA's Navigation Support Office.

GSA (2017) Galileo satellite metadata. https://www.gsc-europa.eu/support-to-developers/galileo-satellite-metadata. Accessed 28 Nov 2017

Guo J, Zhao Q, Geng T, Su X, Liu J (2013) Precise orbit determination for COMPASS IGSO satellites during yaw maneuvers. In:Sun J, Jiao W, Wu H, Shi C (eds) Proc. China satellite navigation conference (CSNC) 2013, vol III. 245:41-53. https://doi.org/10.1007/978-3-642-37407 -4_4

Hauschild A, Steigenberger P, Rodriguez-Solano C (2012) Signal, orbit and attitude analysis of Japan's first QZSS satellite Michibiki. GPS Solutions 16:127-133. https: //doi.org/10.1007/s10291-011-0245-5

Kouba J (2009) A simplified yaw attitude model for eclipsing GPS satellites. GPS Solution 13(1):1-12. https ://doi.org/10.1007/s10291-008-0092-1

Lin X (2018) Satellite Geometry and Attitude Mode of MEO Satellites Developed by SECM, ION GNSS+ 2018

Wang C, Guo J, Zhao Q, Liu J (2018) Yaw attitude modeling for BeiDou I06 and BeiDou-3 satellites. GPS Solutions 22:117. https://doi.org/10.1007/s10291-018-0783-1

Xia F, Ye S, Chen D, Jiang N (2019) Observation of BDS-2 IGSO/MEOs yaw-attitude behavior during eclipse seasons. GPS Solutions 23:71. DOI: https://doi.org/10.1007/s10291-019-0857-8

Xie X (2019) Precise Orbit and Clock Determination for BDS-3 Satellites Using Inter-satellite Link Observations. PhD Dissertation, GNSS Research Center, Wuhan University.

Montenbruck O, Schmid R, Mercier F, Steigenberger P, Noll C, Fatkulin R, Kogure S, Ganeshan A (2015) GNSS satellite geometry and attitude models. Advances in Space Research 56(6):1015-1029. https ://doi.org/10.1016/j.asr.2015.06.019

Zhao Q, Wang C, Guo J, Wang B, Liu J (2018) Precise orbit and clock determination for BeiDou-3 Experimental satellites with yaw attitude analysis. GPS Solution 22:4. https ://doi.org/10.1007/s10291-017-0673-y